\documentclass[superscriptaddress,reprint,amsmath,amssymb,aps,prl,floatfix]{revtex4-2}
\usepackage[T1]{fontenc}
\usepackage[utf8]{inputenc}
\makeatletter
\usepackage{graphicx}
\usepackage{xcolor}
\usepackage{amsmath}
\usepackage{psfrag}
\usepackage[abs]{overpic}
\usepackage{enumitem}
\usepackage{floatrow}
\usepackage{subfigure}
\usepackage{natbib}
\usepackage{braket}
\usepackage{float}
\usepackage{verbatim}
\usepackage{soul}
\usepackage{comment}
\usepackage[normalem]{ulem}
\usepackage{overpic}
\usepackage{babel}
\usepackage[%
colorlinks=true,
urlcolor=blue,
linkcolor=blue,
citecolor=blue
]{hyperref}
\usepackage[shortcuts]{extdash}

\newcommand{\PRLSection}[1]{\emph{#1}.---}

\newcommand{\EJK}[1]{\textcolor{magenta}{#1}}

\makeatother

\begin{document}

\title{Emulating moir\'e materials with quasiperiodic circuit quantum electrodynamics}

\author{T. Herrig}
\email{t-herrig@web.de}
\affiliation{Peter Gr\"unberg Institute, Theoretical Nanoelectronics, Forschungszentrum J\"ulich, D-52425 J\"ulich, Germany}

\author{C. Koliofoti}
\affiliation{Peter Gr\"unberg Institute, Theoretical Nanoelectronics, Forschungszentrum J\"ulich, D-52425 J\"ulich, Germany}

\author{J. H. Pixley}
\affiliation{Department of Physics and Astronomy, Center for Materials Theory, Rutgers University, Piscataway, New Jersey 08854, USA}
\affiliation{Center for Computational Quantum Physics, Flatiron Institute, 162 5th Avenue, New York, NY 10010}

\author{E. J. K\"onig}
\affiliation{Max-Planck Institute for Solid State Research, 70569 Stuttgart, Germany}

\author{R.-P. Riwar}
\affiliation{Peter Gr\"unberg Institute, Theoretical Nanoelectronics, Forschungszentrum J\"ulich, D-52425 J\"ulich, Germany}

\begin{abstract}
Topological bandstructures interfering with moir\'e superstructures give rise to a plethora of emergent phenomena, which are pivotal for correlated insulating and superconducting states of twisttronics materials. While quasiperiodicity was up to now a notion mostly reserved for solid-state materials and cold atoms, we here demonstrate the capacity of conventional superconducting circuits to emulate moir\'e physics in charge space. With two examples, we show that Hofstadter's butterfly and the magic-angle effect, are directly visible in spectroscopic transport measurements. Importantly, these features survive in the presence of harmonic trapping potentials due to parasitic linear capacitances. Our proposed platform benefits from unprecedented tuning capabilities, and  opens the door to probe incommensurate physics in virtually any spatial dimension.
\end{abstract}

\maketitle

\PRLSection{Introduction}Emulating quantum condensed matter systems has emerged as an exciting interdisciplinary frontier that hopes to utilize novel phenomena in solid state systems for technological applications in a range of engineered devices. The available emulator platforms continue to grow and currently range from acoustic and photonic metamaterials, to arrays of microwave resonators, to circuit quantum electrodynamics devices~\cite{Leone2008,Leone_2013,Yokoyama2015TopolABSmultitJJ,Riwar2016,Strambini_2016,Eriksson2017,Meyer:2017aa,Xie:2017aa,Xie:2018aa,Deb:2018aa,Repin:2019aa,Repin_2020,Fatemi_2020,Peyruchat_2020,Klees:2020aa,Klees_2021,Weisbrich_2021_second,Weisbrich_2021_monopoles,Chirolli_2021,Herrig2022CPT,Melo2022}. Several tight binding models have now been realized with a variety of geometries and structures, e.g., exhibiting a band structure with non-trivial topology~\cite{xue2022topological} and higher-order topology~\cite{peterson2018quantized,peterson2020fractional}, or emulated electrons hopping on curved space~\cite{kollar2019hyperbolic}, as well as many-body emulators using ultracold atoms~\cite{RevModPhys.80.885,Celi2014} and superconducting quantum processors~\cite{Cho_2008}.

With the discovery of moir\'e materials in twisted and stacked van der Waals heterostructures composed of graphene mulitlayers~\cite{andrei2021marvels}, transition metal dichalgonides~\cite{mak2022semiconductor}, and cuprate superconductors~\cite{zhao2021emergent}, it is a timely question to understand how to generalize currently available quantum emulators to include moir{\'e} patterns. In this context, it has been understood that these patterns can be implemented by quasiperiodic potentials~\cite{FuPixley2020,CanoWilson2021,GhorashiCano2023} which are either applied externally or arise due to lattice mismatch. This motivated several propositions of ultracold atom emulators to realize the magic-angle effect~\cite{FuPixley2020,ChouPixley2020,SalamonRakshit2020,wangYe2020,MaoSenthil2021,MengZhang2023,LeePixley2022}. More recently, incommensurate effects are taking center stage with the culmination of experimental observations of twisted-graphene quasicrystals~\cite{Ahn2018, UriJarillo2023}. Nonetheless, experiments remain quite challenging due to incorporating specific types of  optical lattices to ensure the combined presence of Dirac points and quasiperiodic potentials.

\begin{figure}
	\centering
    \includegraphics[width=\linewidth]{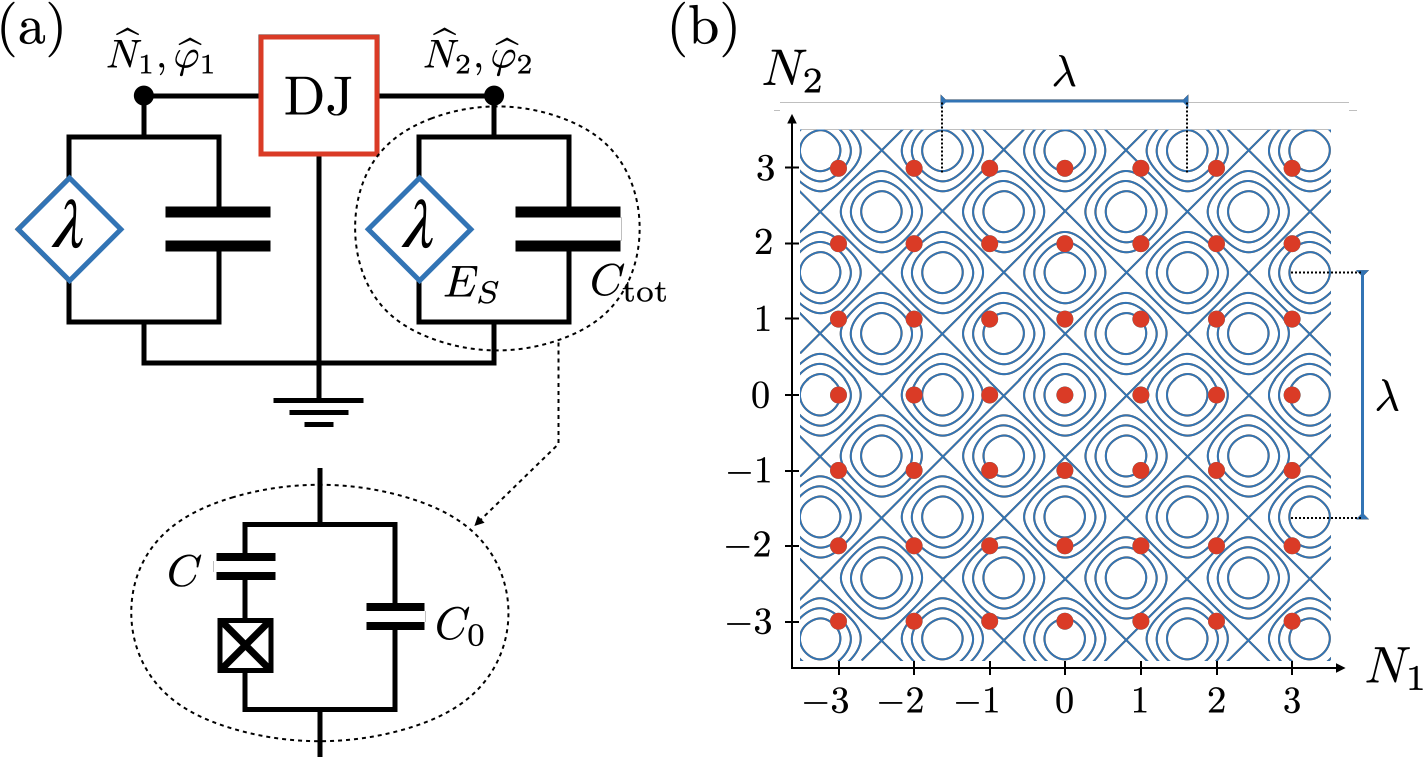}
	
	\caption{Quantum circuit emulating quasiperiodicity in the space of transported charges. 
    (a) 2D-system circuit with a three-terminal Dirac junction (DJ). The two contacts $1$ and $2$ are connected to ground via identical quasiperiodic nonlinear capacitors~\cite{Herrig2023} with quasiperiodicity parameter $\lambda$, and stray capacitances $C_\text{tot}$. The quasiperiodic circuit element is implemented with an auxiliary transmon (inset), where $C_\text{tot} = C + C_0$ and $\lambda = C / C_\text{tot}$. (b) This specific circuit provides a platform to emulate the magic-angle effect in a spin-orbit-coupled tight binding model in a quasiperiodic potential on the square lattice. The red dots mark the discrete eigenvalues of the charge states on the two nodes, $N_1$ and $N_2$, forming the tight-binding lattice. The blue contour plot depicts the quasiperiodic potential (with periodicity $\lambda$). For nonrational $\lambda$ the potential is incommensurable with respect to the quantized charges $N_{1,2}$.
	}
	\label{fig:circuit-TBG}
\end{figure}

In this letter, we show that the conventional toolbox of circuit quantum electrodynamics (cQED) (i.e. non-topological Josephson junctions, capacitances and inductances) allows to create and measure moir\'e effects in the transport degrees of freedom of a quantum circuit (see Fig.~\ref{fig:circuit-TBG}). While we explicitly illustrate the principle with the emulation of an analog Hofstadter butterfly and the magic-angle effect, note that our approach harbors the potential to effectively realize the physics of virtually any dimension~\cite{Riwar2016,Weisbrich_2021_second}, since the lattice is defined in the number of transported charges instead of the actual position space. Moreover, the generation of a nontrivial band structure occurs in a circuit element distinct from the one giving rise to the quasiperiodic potential, allowing for an exceptional degree of tunability. In particular, there exists the possibility of an in-situ control of the quasiperiodicity parameter by tuning the circuit capacitances~\cite{riwar2023discrete}. 
Finally, the proposed circuits implement single-particle physics only. While, beyond doubt, the correlated many-body phases of twisted magic-angle samples are of exceptional excitement, they also wash out and bury interesting single-particle physics, including (multi\=/)fractal features~\cite{FuPixley2020,gonccalves2021incommensurability} occurring when the twist angle is incommensurate. 

Note that our work builds on recently proposed quasiperiodic capacitive elements~\cite{Herrig2023} introduced as a blue-print to realize cQED emulators of models for quasiperiodic Anderson localization. The presence of parasitic linear capacitances however prevented an explicit link between these fields to be made. Therefore, a central point of our endeavor concerns the impact of said parasitic capacitances, leading to a harmonic trapping potential, similar to the harmonic trapping potential conventionally used  in cold atom experiments. As we explicitly demonstrate here, the known features in the density of states for both Hofstadter butterfly and magic angle effect in the absence of harmonic traps have a one-to-one correspondence in the density of states with traps. This makes the observation of these features amenable to standard ac spectroscopy tools, readily available in quantum circuit hardware. Furthermore, given the omnipresence of harmonic traps in other physical platforms, this surprising result is relevant beyond the here considered quantum circuit context.

\PRLSection{Quasiperiodic circuit QED}In our earlier work~\cite{Herrig2023} we have introduced the quasiperiodic nonlinear capacitor (QPNC) as a new circuit element realizing an energy contribution of the form
\begin{equation}
    \widehat{H}_{\mathrm{QPNC}} = - E_S \cos\left(2\pi \lambda \widehat{N}\right) + 2E_C \widehat{N}^2\ ,
\label{eqn:HQPNC}
\end{equation}
where \(\widehat{N}\) is the number of Cooper pairs separated across the capacitor. This element can be realized with conventional superconducting materials. The nonlinear term arises from a capacitively coupled auxiliary transmon~\cite{Herrig2023} [see also inset in Fig.~\ref{fig:circuit-TBG}(a)], where $E_S$ corresponds to the quantum phase slip amplitude within the transmon~\cite{Koch_2007}. This energy scale can be controlled in a time-dependent fashion by tuning the critical current of the transmon's Josephson junction. The quasiperiodicity parameter \(\lambda\) corresponds to the  ratio of the capacitive coupling of the transmon, $C$, with respect to the total capacitance, $C_\text{tot}$. The latter defines the charging energy \(E_C=2e^2/C_\text{tot}\) of the accompanying (parasitic) linear capacitance. Therefore, \(\lambda\) is in general a real, incommensurate parameter and renders the charging energy quasiperiodic with respect to the discrete charge observable \(\widehat{N}\). In~\cite{Herrig2023} we explored the tunability of the parameter \(\lambda\) and the energy scales \(E_S\) and \(E_C\), and showed in particular how to tune the device into a regime of ${\lambda^2}E_S > E_C$, by means of engineering effective attractive interactions, an experimentally verified principle in quantum dots~\cite{Little_1964,Hamo_2016,Placke_2018}. This will be the regime of interest in what follows. Since $\lambda$ depends on capacitances, measuring features as a function of $\lambda$ is a little more sophisticated. One could either fabricate a series of devices with different geometry on the same chip to tailor the electrostatic properties, or $\lambda$ may be amenable to time-dependent tuning within a single device, due to a recently proposed tunability of circuit capacitances~\cite{riwar2023discrete}.

In this work, we explore the possibility to embed the QPNC into larger circuits to create Hamiltonians with a nontrivial density of states (DOS), and how to measure the DOS. In particular, our focus is on investigating the role of the parasitic linear capacitive energy term $\sim E_C$, which will prove to be pivotal, in spite of being small. All considered model systems can be described by
\begin{equation}\label{eq:general-Hamiltonian}
    \widehat{H}_{d} = \widehat{H}_{J} \left(\widehat{\varphi}_{1}, \ldots \widehat{\varphi}_{d}\right) + \sum_{j=1}^{d} \widehat{H}_{\mathrm{QPNC}}^j
\end{equation}
where each \(\widehat{H}_{\mathrm{QPNC}}^j\) is given by Eq.~\eqref{eqn:HQPNC} after inserting the respective island charge \(\widehat{N}_j\). \(\widehat{H}_{J}\) represents a generic Hamiltonian describing a multi-terminal junction, connecting \(d+1\) superconducting contacts (for an example with \(d=2\), see Fig.~\ref{fig:circuit-TBG}). One contact is put to ground, such that its superconducting phase can be fixed to zero. There then remain $d$ islands, each of which is connected with an individual QPNC. The islands' degrees of freedom are the canonically conjugate phase and charge operators \(\bigl[\widehat{\varphi}_{j}, \widehat{N}_{k}\bigr] = i\delta_{jk}\). 

The charge space spanned by the eigenstates of all $\widehat{N}_j$ can thus be considered as the equivalent of a $d$-dimensional lattice, where the junction $\widehat{H}_J$ encodes tunneling between lattice points and (optionally) intrinsic degrees of freedom which act as a pseudo spin. The QPNC provides a quasiperiodic potential ($\sim E_S$) on top of a harmonic trap ($\sim E_C$). For simplicity, we assume that $E_S$ and $E_C$ are the same for each island. 

Each charge island is subject to an induced offset charge. Note that due to the details of its implementation the QPNC is affected by two independent sources of offset charges~\cite{Herrig2023} for the quasiperiodic (\(\sim E_S\)) and for the quadratic term (\(\sim E_C\)). These independent offset charges are included in Eq.~\eqref{eq:general-Hamiltonian} by shifting \(\widehat{N}_{j} \longrightarrow \widehat{N}_{j} + N_{g,j}^{\alpha}\) where $\alpha = C,S$, respectively. 

Our main focus is the analysis of the DOS of $\widehat{H}_{d}$,
\begin{equation}
   \rho(E) = \frac{1}{D} \sum_n \delta(E - \epsilon_n)
\end{equation}
where $\epsilon_n$ are the eigenvalues of $\widehat{H}_{d}$ and $D$ is the size of the Hilbert space. We hone in on two example systems. We study a 1D system, where the only island is connected to ground with an ordinary Josephson junction, and a 2D system with a Dirac junction, i.e., a three-terminal junction showing a Dirac spectrum [see Fig.~\ref{fig:circuit-TBG}(a)]. For $E_C=0$, the 1D system implements the Aubry-Andr\'e model~\cite{AubryAndre,Harper_1955} (whose spectrum hosts Hofstadter's butterfly~\cite{Hofstadter1976}), whereas the 2D system simulates a spin-orbit-coupled tight binding model, subject to an additional quasiperiodic potential, which at weak to moderate quasiperiodic strength gives rise to the same phenomena as the magic-angle in twisted bilayer graphene, including flat and isolated minibands, multiple magic-angles, and emergent Dirac excitations on the moir\'e scale~\cite{FuPixley2020,ChouPixley2020,YiPixley-2022,gonccalves2021incommensurability} (but Anderson-localizes at larger values of the potential).

Some general remarks are in order. First, there is a crucial difference between the here considered circuit implementation, compared to many-body solid state or cold-atomic systems. If the lattice points actually referred to positions of particles, then there would be a meaningful way to extend the description to a many-particle wave function. The lattice `positions' in $\widehat{H}_{d}$ on the other hand (see Fig.~\ref{fig:circuit-TBG}) already \textit{are} in second quantized form, as the observables $\widehat{N}_j$ count the number of charges on island $j$. Therefore, the ground state of $\widehat{H}_{d}$, $\epsilon_0$, is the many-body ground state, and there is no meaningful way to emulate quantities like, e.g., a chemical potential. This is an important, and as a matter of fact, advantageous feature. Namely, by means of probing the ac response of a circuit realizing $\widehat{H}_{d}$, one is capable of directly measuring the DOS that would correspond to the single-particle DOS if $\widehat{H}_{d}$ referred to positions of a particle instead of the circuit's charge configuration~\footnote{As already hinted at in the introduction, an ac driving of solid state systems or cold atom gases would in general generate a more complicated response involving many-body excitations and interactions.}. It is in this sense, that we refer to $\rho$ as a single-particle DOS. 

The second important remark concerns the role of the parasitic linear capacitances, $\sim E_C$. In the existing literature, the DOS for the above models has been most widely studied in a condensed matter context, where there is no harmonic trapping potential, defining the un\-trapped model $\widehat{H}_0 \equiv \widehat{H}_{d}(E_C=0)$. We here find that while the DOS of $\widehat{H}_0$, $\rho_0(E)$, is related to $\rho(E)$, the connection is nontrivial. In particular, $\rho(E)$ for the system with finite $E_C$ does \textit{not} simply approach $\rho_0(E)$ (the `untrapped' system) when going to the limit $E_C\rightarrow 0$~\footnote{While $E_C$ can be tuned to very small values (see Ref.~\cite{Herrig2023}), it is practically impossible, respectively physically meaningless, to have exactly zero charging energy \(E_C=0\).}. This is due to the fact that there is no perturbative limit for the trapping energy as the potential strength diverges sufficiently far away from the origin. Instead, as $E_C \rightarrow 0$, we find the convolution formula~\cite{suppl}
\begin{equation}\label{eq_DOS_trap_vs_no_trap}
    \rho(E) \approx \int^E_{E_{0}} dE^\prime\left(E - E^\prime\right)^{\frac{d}{2} - 1} \rho_0\left(E^\prime \right) \stackrel{d=2}{\rightarrow} \partial_E \rho = \rho_0,
\end{equation}
where $E_0$ is the bottom of the band. Consequently, the DOS gets distorted in a way characteristic of dimensionality $d$. For instance, a delta peak in $\rho_0(E)$ morphs either into a van-Hove singularity with algebraic tail $\sim 1/\sqrt{E}$ (for $d=1$) or into a Heaviside step function (for $d=2$) in $\rho(E)$. More generally, in $d=1$ peak features survive, whereas in $d=2$, peak features translate into staircase-like features, as reflected in the simple relationship given in the second equality in Eq.~\eqref{eq_DOS_trap_vs_no_trap}. We now proceed with numerical calculations~\cite{suppl} of the above introduced explicit models with $d=1,2$.

\PRLSection{Aubry-Andr\'e model and Hofstadter Butterfly}We begin by exploring quasiperiodicity in a simple 1D system in conjunction with a harmonic trap~\cite{Modugno2009}. While aspects of this problem have previously been studied with ultracold atoms~\cite{RoatiInguscio2008,LuschenBloch2018,WangJia2022}, as well as acoustic~\cite{NiKhanikaev2019}, photonic \cite{LahiniSilberberg2009,KrausZilberberg2012}, and polaritonic~\cite{GoblotZilberberg2020} emulators, quantum circuits present another, extremely compact platform, where circuit-specific observables allow a direct measurement of the characteristic Hofstadter DOS.

\begin{figure}
	\centering
	\includegraphics[width=\linewidth]{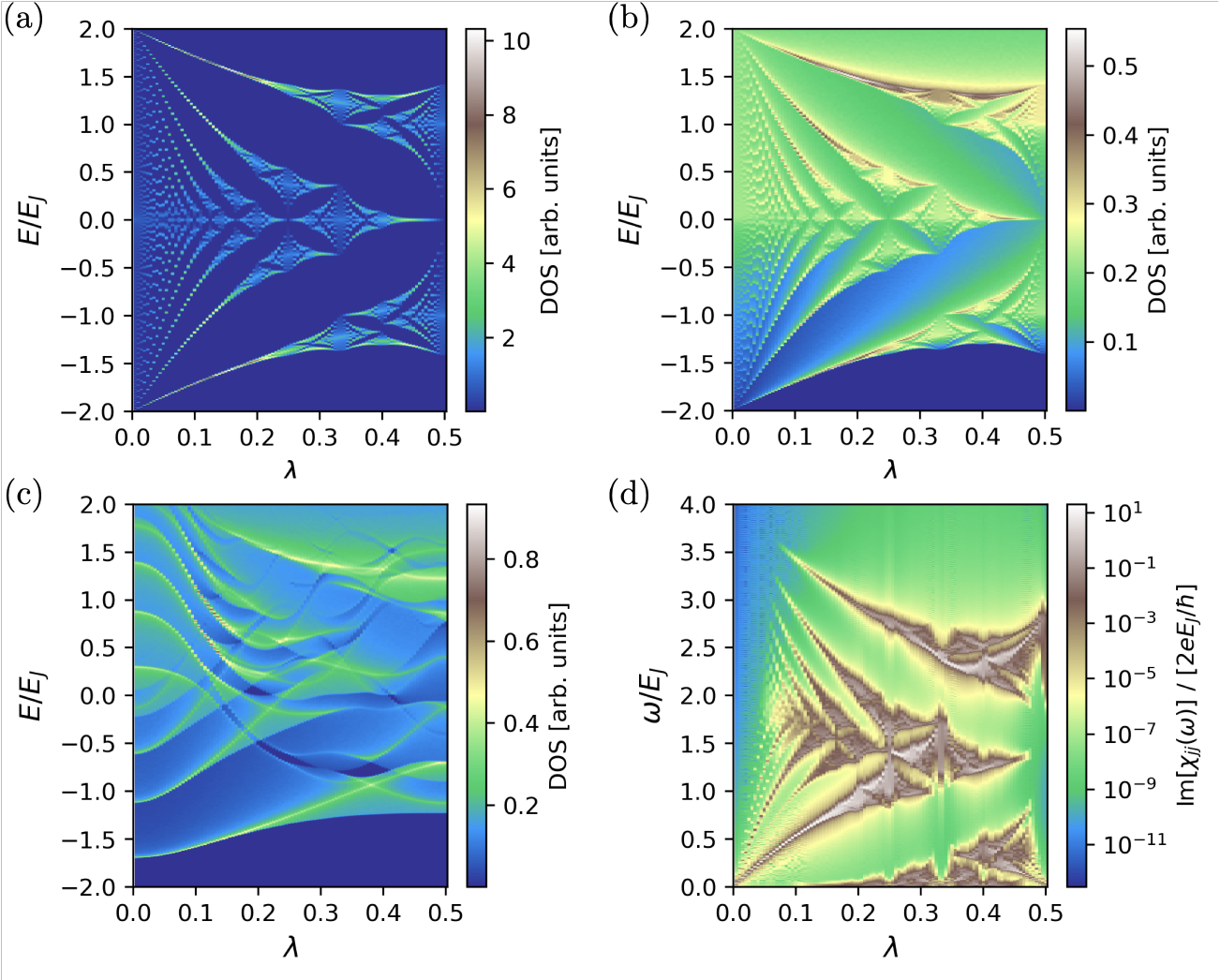}
	
	\caption{Spectral functions of Hofstadter-butterfly-emulator. We consider the 1D system with \(E_J = E_S\), for (a) \(E_C = 0\) (the original Hofstadter butterfly), (b) \(E_C \sim 10^{-4} E_J\), and (c) \(E_C = 10^{-1} E_J\). (d) The auto-correlation of the current response on a logarithmic scale with \(E_C \sim 10^{-4} E_J\) resembles the density of states shown in (b). The original Hofstadter butterfly (a) was computed in a discretized phase space with 600 lattice points, while for finite $E_C$ the computations were done on a charge lattice of size (b, d) 271, respectively (c) 11. The expansion orders of the kernel polynomial method are, respectively, (a) 651, (b, d) 1474, and (c) 1822.
	}
\label{fig:Hofstadter_butterflies}
\end{figure}

Consider \(\widehat{H}_{d=1}\) from Eq.~\eqref{eq:general-Hamiltonian} with the central junction being a simple Josephson junction, $\widehat{H}_{J}^{d=1} = -E_{J} \cos\left(\widehat{\varphi}_{1}\right)$, where \(E_J = E_S\).
For \(E_C=0\) this is the Aubry-André model, which produces the famous Hofstadter butterfly (HB) pattern~\cite{Hofstadter1976}; see Fig.~\ref{fig:Hofstadter_butterflies}(a).  In accordance with Eq.~\eqref{eq_DOS_trap_vs_no_trap}, the influence of the finite trapping term $E_C\neq 0$ is nontrivial. Energy gaps in the pure HB are now filled with a background density of states, due to the already mentioned algebraic tails $\sim 1/\sqrt{E}$; see Figs.~\ref{fig:Hofstadter_butterflies}(b, c). Nevertheless, for \(E_C / E_J \sim 10^{-4}\), panel (b), the HB peaks are clearly visible. When going to values close to \(E_C / E_J \sim 10^{-1}\), Eq.~\eqref{eq_DOS_trap_vs_no_trap} starts to lose its validity, see Fig.~\ref{fig:Hofstadter_butterflies}(c). Note that \(E_C / E_J \sim 10^{-2}\) corresponds to the regime where transmon devices are typically operated.

The DOS of a quantum circuit can be straightforwardly measured via ac spectroscopy in various manners~\cite{Blais2021cQED}, similar in spirit to Refs.~\cite{Fuechsle:2009aa,Bretheau:2013ab,Bretheau:2013aa,Novikov2013,Woerkom:2017aa,Shulga:2017aa,Proutski:2019aa,Tosi:2019aa,Bargerbos:2022aa,willsch2023observation}. For concreteness, we here explicitly study a specific measurement of the DOS via a current response due to an external flux. The standard linear response to a time-dependent magnetic flux \(\varphi_{j} \rightarrow \varphi_{j} + \varphi_{\text{ext}}^{j}\left(t\right) \theta\left(t-t_{0}\right)\) yields the autocorrelated response function~\cite{suppl} 
\begin{align}\label{eq:autocorrelator}
	\operatorname{Im} \left[\chi_{jj} \left(\omega>0\right)\right] = \pi \left|\left\langle \epsilon_{0}\right| \widehat{I}_{j} \left|\epsilon_{0} + \omega\right\rangle \right|^{2} \rho\left(\epsilon_{0} + \omega\right),
\end{align}
with the current operator \(\widehat{I}_{j} = -2e\, i\bigl[\widehat{H}_{J}, \widehat{N}_{j}\bigr]\), eigenenergies \(\epsilon_{n}\), and eigenstates \(|\epsilon_{n}\rangle\).
We show this response in Fig.~\ref{fig:Hofstadter_butterflies}(d) with a logarithmic color scale in units of \(2e E_{J} / \hbar\). We observe that while the HB pattern still persists, it is distorted with respect to Figs.~\ref{fig:Hofstadter_butterflies}(a--c). This is simply because for the response function the $\lambda$-dependent ground state energy is automatically set to zero as a reference point.

\PRLSection{Magic-Angle effect in 2D Dirac system}We now move on to consider a three-terminal junction [$d=2$ in Eq.~\eqref{eq:general-Hamiltonian}] with a generic two-dimensional Dirac structure
\begin{equation}
    \widehat{H}_{J}^{D} = E_{DJ}
    \begin{pmatrix}
        0 & \sin\left(\widehat{\varphi}_{1}\right) + i\sin\left(\widehat{\varphi}_{2}\right)\\
        \sin\left(\widehat{\varphi}_{1}\right) - i\sin\left(\widehat{\varphi}_{2}\right) & 0
    \end{pmatrix},
\end{equation}
see Fig.~\ref{fig:circuit-TBG}.
Such a Dirac junction can be realized, e.g., with four-terminal Weyl circuits~\cite{Riwar2016,Peyruchat_2020,Fatemi_2020}, by closing two contacts by a flux-threaded loop, and tuning the external flux to the degeneracy point. While in this construction, the Dirac points would not be topologically protected, residual side-effects of a fluctuating minigap due to flux noise could likely be mitigated by increasing the loop area (at the cost of a non-negligible loop inductance in the limit of very large loops). Dirac physics have also been predicted in several other conventional circuits (consisting of regular Josephson junctions), either in
a mixed 2D phase-charge space~\cite{Herrig2023,Chirolli_2021} or in a pure
charge-charge space~\cite{Herrig2023,Chirolli_2021}. We expect that charge and phase degrees of freedom in these proposals can be swapped by exploiting the duality between Josephson and quantum phase slip junctions~\cite{Ulrich_2016}.

\begin{figure}
	\centering
	\includegraphics[width=\linewidth]{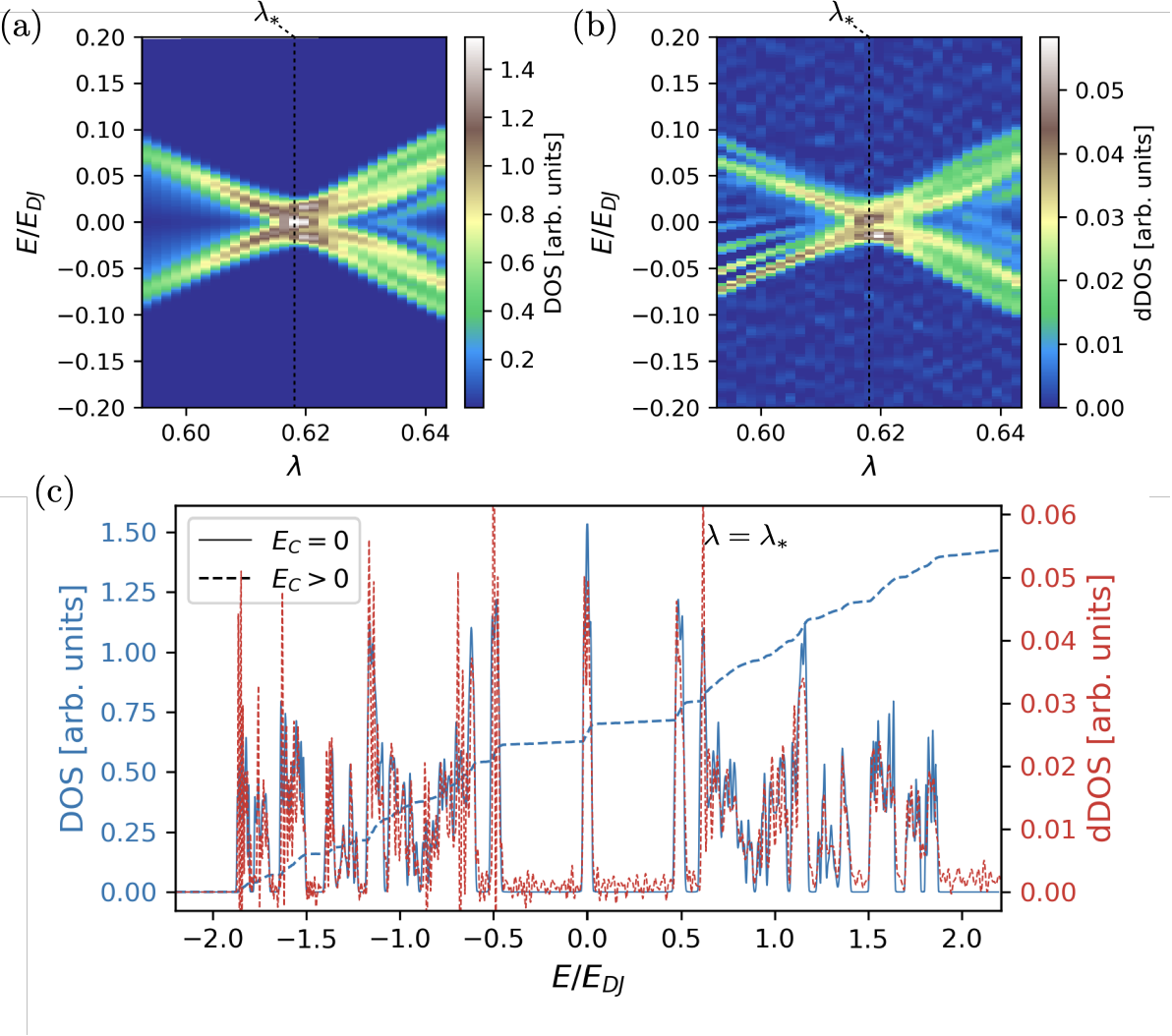}
	
	\caption{Spectral signatures of the magic-angle emulator. Here, with \(E_S/E_{DJ} = 0.541\) a magic-angle appears at \(\lambda = \lambda_* \approx 0.62\). 
    (a) The density of states (DOS) for \(E_C = 0\) shows the characteristic magic-angle effect. 
    (b) For a finite \(E_C \sim 10^{-4} E_{DJ}\) the same characteristic pattern emerges in the differential DOS (dDOS), \(\partial_E \rho\left(E\right)\). Note that some weakly negative values have been cut to match the color scale of panel (a)~\cite{suppl}. 
    (c) Comparison of the DOS for $E_C/E_{DJ}=0,10^{-4}$ (blue solid and blue dashed) and dDOS for $E_C/E_{DJ}=10^{-4}$ (red dashed), all at $\lambda = \lambda_*$. The DOS for finite $E_C$ has been scaled up by a factor of 40 to allow for a direct comparison. 
    All these spectra were computed in a discretized phase space with $610^2$ lattice points, while the expansion orders of the kernel polynomial method are, respectively, (a) 1430, (b) 4019, and (c) 1430 ($E_C = 0$) and 4019 ($E_C > 0$).
    \label{fig:magic_angle}
	}
\end{figure}

For \(E_C = 0\), it was previously demonstrated~\cite{FuPixley2020} that $\widehat{H}_{d=2}^D=\widehat{H}_J^D+\sum_{j=1}^2\widehat{H}^j_{\mathrm{QPNC}}$ exhibits magic-angle physics in the same universal fashion as twisted bilayer graphene~\cite{bistritzer2011moire,FuPixley2020,gonccalves2021incommensurability} with the characteristic flat bands~\cite{YiPixley-2022} and non-zero DOS at zero energy for critical values of $E_S$ and $\lambda$, see Fig.~\ref{fig:magic_angle}(a). Consistently with Eq.~\eqref{eq_DOS_trap_vs_no_trap} (and subsequent discussion) we here numerically demonstrate that the same transition can also be observed in the presence of a finite \(E_C > 0\). The characteristic magic-angle structure of the DOS now reveals itself in the form of steps surrounded by plateaus of constant DOS (or as peaks in the differential DOS, referred to as dDOS) at the precise values of energy, where the system without trap exhibits flat band peaks, see Figs.~\ref{fig:magic_angle}(b, c). This result numerically confirms the convolution formula, Eq.~\eqref{eq_DOS_trap_vs_no_trap}, for $d=2$. In analogy with the previous section, one can measure this feature by means of an ac response of the circuit, here in particular with a form of differential response $\partial_\omega\chi_{jj}$. Once again, if the device is prepared in the ground state [roughly at $-2E_{DJ}$ in Fig.~\ref{fig:magic_angle}(c)] the magic angle peak will be visible at a driving frequency of $\sim 2 E_{DJ}$. This finding constitutes one of our main results: the emulation of magic-angle physics in superconducting circuits requiring a relatively small number of conventional circuit elements. Finally, we insist that harmonic traps are a feature transcending the narrow context of superconducting circuits and our specific realization of QPNCs~\cite{Herrig2023}, such that the demonstrated survival of the magic angle feature in the presence of such traps [underlined by Eq.~\eqref{eq_DOS_trap_vs_no_trap}] is expected to be of importance also for many other platforms.

\PRLSection{Conclusion and outlook}We demonstrate that conventional circuit elements can reproduce the DOS from a wide variety of solid state moir\'e materials, such as the Hofstadter butterfly and the magic angle effect. We show both analytically and numerically that distortions of the DOS due to harmonic traps (parasitic linear capacitances) do not impede the observation of characteristic features---a finding relevant beyond the superconducting circuit context. We further show that straightforward ac current responses directly probe the DOS. Our results prepare the ground for a multitude of future research directions, such as the inclusion of effective many-body features by increasing the number of circuit degrees of freedom. The main challenge for such a generalization is likely the imitation of bosonic (fermionic) exchange statistics, exactly because the wave function is already embedded in a many-body Hilbert space, and as such does not naturally have the required (anti\=/)symmetry in charge space. Another nontrivial extension is to use the circuit-specific flexibility to emulate topological materials with $d>3$ (see, e.g., Ref.~\cite{Weisbrich_2021_second} for a junction emulating a topological material with $d=4$ exhibiting a nonzero second Chern number), and study the so far poorly understood interplay between moir\'e patterns and topological band structures beyond the usual limitation of three (spatial) dimensions.

\begin{acknowledgments}
	\PRLSection{Acknowledgments}T.H. and R.P.R. gratefully acknowledge computing time on the supercomputer JURECA~\cite{JURECA} at Forschungszentrum Jülich under grant no. JIFF23. This work has been funded by the German Federal Ministry of Education and Research within the funding program Photonic Research Germany under the contract number 13N14891.
	J.H.P.\ is partially supported by the Air Force Office of Scientific Research under Grant No.~FA9550-20-1-0136 and the Alfred P.\ Sloan Foundation through a Sloan Research Fellowship. The Flatiron Institute is a division of the Simons Foundation.
\end{acknowledgments}

\bibliography{magic_angle}

\clearpage

\setcounter{equation}{0}
\setcounter{figure}{0}
\setcounter{section}{0}
\setcounter{table}{0}
\setcounter{page}{1}
\makeatletter
\renewcommand{\theequation}{S\arabic{equation}}
\renewcommand{\thesection}{S\arabic{section}}
\renewcommand{\thefigure}{S\arabic{figure}}
\renewcommand{\thepage}{S\arabic{page}}

\begin{widetext}
\begin{center}
Supplementary materials on \\
\textbf{``Emulating the magic-angle effect in quasiperiodic circuit quantum electrodynamics''}\\
T.~Herrig, C.~Koliofoti, J.~H.~Pixley, E.~J.~K\"onig, and R.-P.~Riwar
\end{center}
\end{widetext}

\section{Numerical methods}

Let us begin with some general remarks. We use Python to implement the kernel polynomial method~\cite{WeisseFehske2006} which allows to approximate and compute spectral functions like a DOS or a correlation function without diagonalization of the Hamiltonian. It is based on a Chebyshev expansion while applying the Jackson kernel to avoid Gibbs oscillations. In order to compute specific eigenenergies or \=/states like the ground state, we made use of the Lanczos algorithm. Both of these methods scale linearly with the system size and take advantage of sparse matrices.

\subsection{Approximation of the Hilbert space}

When numerically implementing Hamiltonians of the here considered form [see Eq.~(2)
] whose Hilbert space is infinite we have to approximate them such that they only live in a finite Hilbert space. Considering that the conjugated variables are pairs of a discrete charge and a compact phase, we found the following two approaches useful.

That is, either we introduce a cutoff in charge space or we discretize the phase space. The first approach is only useful with a finite charging energy $E_C > 0$ since the corresponding energy parabola in charge space separates the charge values energetically such that we find a good approximation for the lower part of the spectrum for a chosen cutoff $|N| < N_\text{co}$ (without altering form of the Hamiltonian).
The second approach, on the other hand, is especially useful for $E_C = 0$ where the above approximation breaks down. Here, the quasiperiodic term with $\lambda = p / q$ (this approach only works with commensurate values of $\lambda$ but we can approach incommensurate values with $q \gg 1$) represents a hopping term translating the phase by \(2\pi\lambda\). This defines a grid in phase space of \(\varphi_k = 2\pi k/q\) which we use to discretize the phase space. The values between these points are then covered by including an offset value $0 \leq \Delta\varphi < 2\pi / q$ as a new parameter analogous to an offset charge, which technically also has to be integrated out. In praxis, however, increasing $q$ suppresses any dependence on $\Delta\varphi$.

\begin{figure}
	\centering
    \includegraphics[width=\linewidth]{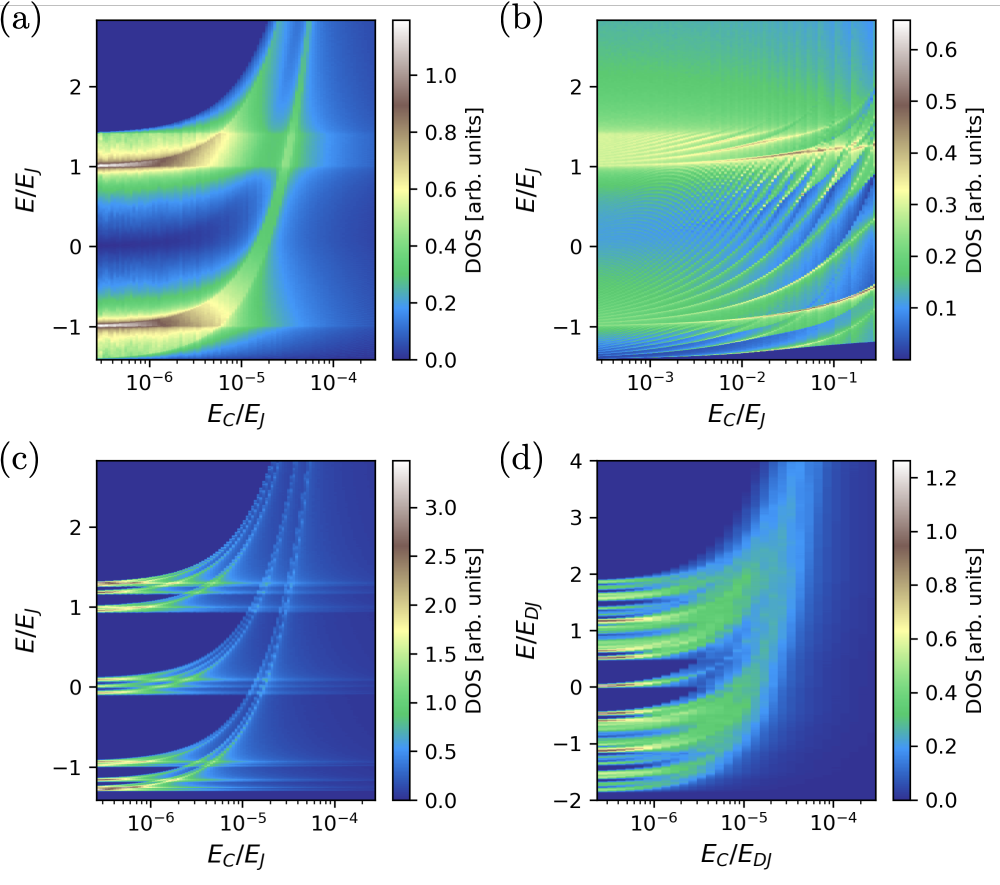}

	\caption{Smooth introduction of the harmonic trap. (a, c, d) The trap is approximated by a cosine potential in the low \(E_C / E_J\) regime showing the transition of a negligible to a quadratic \(E_C\) contribution. (b) Introducing a charge cutoff for higher values of \(E_C / E_J\), the spectrum becomes more complex. The quasiperiodicity parameter is chosen to be (a, b) \(\lambda = 1/2\), and (c, d) \(\lambda \approx 0.62\), while (a--c) \(E_S = E_J\) or, respectively, (d) \(E_S / E_{DJ} = 0.541\). The spectra were computed in a discretized phase space with (a) 600, (c) 610, or (d, e) \(610^2\) lattice points or (b) on a charge lattice with the size ranging between 7--191 (needing a larger lattice for smaller \(E_C\) values). The expansion orders of the kernel polynomial method respectively range between (a) 463--3756, (b) 1371--2001, (c) 474--3767, and (d) 1443--14614.
	}
    \label{fig:suppl_EC_trans}
\end{figure}

\subsection{Cosine approximation of the harmonic trap}

In the second approach of a discretized phase space we cannot implement an exact charge parabola $\sim N^2$ since this term corresponds to infinitesimal shifts in the phase space. However, we can introduce an approximation of a term creating the smallest shifts possible
\begin{equation} \label{eq:QPCapacitance}
    (\widehat{N}_j + N_{g,j}^C)^2 \longrightarrow \frac{q^{2}}{2 \pi^{2}} \left[1 - \cos\left(2\pi \frac{\widehat{N}_{j} + N_{g,j}^C}{q}\right)\right],
\end{equation}
that is, a nearest neighbor hopping term in the discretized phase space $\sum_k \left|\varphi_{k+1}\right\rangle \left\langle \varphi_k\right| + \text{h.c.}$, whose cosine form approximates a charge parabola for increasing values of $q$. This, again, gives us a good approximation for the lower part of the spectrum. However, it also provides the possibility to introduce a finite value of $E_C$ in a smooth manner, starting from $E_C = 0$.
This transition of the DOS is shown in Figs.~\ref{fig:suppl_EC_trans}(a, c, d) for various cases, where we fixed $\lambda$ and $q$ while increasing $E_C / E_S$ from $q^2 E_C / \pi^2 \ll E_S$ to $q^2 E_C / \pi^2 \gg E_S$. This represents the transition into the limit of low but finite harmonic trap. Beyond that, Fig.~\ref{fig:suppl_EC_trans}(b) demonstrates the effect of larger $E_C$ values.

\subsection{\texorpdfstring{$E_C$}{E\textunderscore C}-dependent adjustment of numerical parameters}

For the plots of Fig.~\ref{fig:suppl_EC_trans} where we tuned $E_C$ over multiple orders of magnitude, we adjusted the expansion order of the kernel polynomial method $N_{\text{KPM}}$ with increasing values of $E_C / E_J$ to scale with the size of the spectrum $N_{\text{KPM}} \sim E_\text{max} - E_\text{min}$. Moreover the charge cutoff $N_\text{co}$ in panel (b) is also adjusted according to $N_\text{co} = \left\lceil \sqrt{5 E_J / 2 E_C}\right\rceil$, where $\left\lceil x\right\rceil$ maps $x$ to the least integer greater than or equal to $x$.

\begin{figure}
	\centering
    \includegraphics[width=\linewidth]{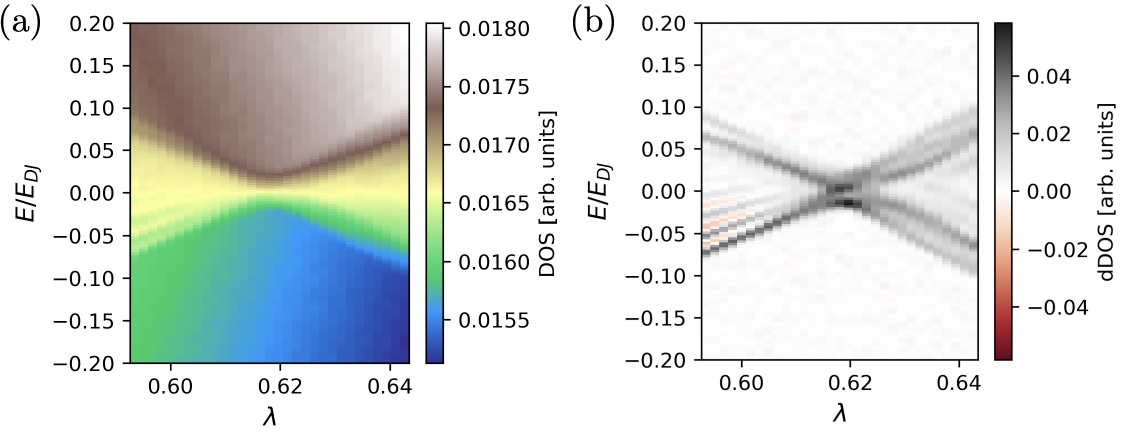}

	\caption{Staircase-like magic-angle signature in the trapped model.
    (a) In the presence of the harmonic trap, we find a density of states (DOS) which is characterized by steps instead of peaks but still incorporates the magic angle effect. (b) The energy derivative of the DOS (dDOS) reveals the characteristic magic-angle signature from the DOS of the untrapped model. The quasiperiodicity parameter is \(\lambda \approx 0.62\), while \(E_S / E_{DJ} = 0.541\). The spectra were computed in a discretized phase space with \(610^2\) lattice points. The expansion order of the kernel polynomial method is 4019.
	}
    \label{fig:suppl_magic_angle}
\end{figure}

\subsection{Magic-angle signature in the trapped model}

In the presence of the harmonic trap, the characteristic spectral signature of the magic-angle effect gets incorporated into the energy derivative of the DOS (dDOS) while the DOS itself shows the same pattern in staircase-like manner. That is, peaks in the DOS of the untrapped system turn into steps while empty density turns into a plateau of constant DOS in the trapped model, see Fig.~\ref{fig:suppl_magic_angle}. The computed dDOS shows weak negative values which we fully present here. In the main text, however, we cut these values off in order to match the color map of the computation for the untrapped model to render the two more comparable; see Fig.~3
(a, b).

\section{Density of states at finite trapping potential}

We here derive the convolution formula given in Eq.~(4) 
in the main text. To this end, we start from a generic $d$-dimensional Hamiltonian with harmonic
trapping as defined in Eq.~(2)
. As in the main text, we define the untrapped Hamiltonian as $\widehat{H}_0=\widehat{H}_d(E_C=0)$. The full Hamiltonian can thus be decomposed into $\widehat{H}_d=\widehat{H}_0+\widehat{H}_\text{trap}$, where the last term simply encompasses the trapping energy term, $\widehat{H}_\text{trap}=2E_C\sum_j \bigl(\widehat{N}_j + N_{g,j}^C\bigr)^2$.
While Eq.~(4) 
is valid in the limit $E_C\rightarrow 0$, where the spectrum no longer depends on the offset charges $N_{g,j}^C$, it actually turns out that the offset charges play an essential role in the proof, as we show in what follows.

We define the DOS as in the main text,
\begin{equation}\label{eq_GR_and_rho}
    \rho\left(E\right) = -\frac{1}{\pi} \operatorname{Im}\left[G^{R}\left(E\right)\right]
\end{equation}
where $G^{R}$ represents the retarded Green's function
\begin{equation}
    G^{R}\left(\omega\right) = \operatorname{tr}\left[\frac{1}{\widehat{H}_d - \omega + i0^+}\right].
\end{equation}
In fact, the proof works interchangeably for $G^{R}$ and $\rho$. Since the Green's function is the more general of the two quantities, we show it explicitly for $G^R$. The corresponding Green's function for $\widehat{H}_{0}$ is
\begin{equation}
    G_{0}^{R}\left(\omega\right) = \operatorname{tr}\left[\frac{1}{\widehat{H}_{0} - \omega + i0^+}\right].
\end{equation}
We now make use of the hypothesis stated in the main text, that for $E_{C}\rightarrow0$,
the spectrum of $H$ does not depend on the offset charges, that is
$\operatorname{spec}\left[\widehat{H}_d\bigl(N_{g,1}^C, N_{g,2}^C, \ldots, N_{g,d}^C\bigr)\right] \approx \operatorname{spec}\left[\widehat{H}_d\left(0, 0, \ldots, 0\right)\right]$.
\textit{Crucially}, note that this statement involves a gauge fixing.
In principle, we might apply a gauge transformation on the full Hamiltonian
\begin{equation}
    \widehat{H}_d^{\prime} = \widehat{H}_d + f\left(N_{g,1}^C, N_{g,2}^C, \ldots, N_{g,d}^C\right),
\end{equation}
where $f$ is an arbitrary scalar function. Both $\widehat{H}_d^\prime$ and $\widehat{H}_d$ predict the exact same physics. But we note that our above
statement about the invariance of the spectrum on $N_{g,j}$
is true if, and only if, $f\equiv\text{const.}$ (for simplicity, $f\equiv 0$ in what follows) for all values of
$N_{g,j}$. Consequently, for the system in
the gauge $\widehat{H}_d$ (and \textit{only} in this gauge), the statement
\begin{equation}\label{eq_gauge_equality}
    G^{R}\left(\omega\right) = \frac{1}{\mathcal{N}} \int dN_{g,1}^C \int dN_{g,2}^C \ldots \int dN_{g,d}^{C} G^{R}\left(\omega\right)
\end{equation}
must be true. The variable $\mathcal{N}$ is a suitably chosen (but irrelevant) normalization constant. By means of Eq.~\eqref{eq_gauge_equality} we can fully appreciate the importance of the gauge fixing. Had we not chosen the above gauge, we would integrate the density of states on the right-hand side of the equation by an arbitrary and unphysical energy reference point which fluctuates with the offset charges. This would be in contradiction with the basic fact that only energy differences are measurable (e.g., from ground to excited states).

\begin{widetext}

On a side note, we emphasize the importance of performing the
trace in $G^{R}$. We cannot make the
same statement on the operator object itself (prior to taking the
trace). That is, in general
\begin{equation}
    \frac{1}{\widehat{H}_d - \omega + i0^+} \neq \frac{1}{\mathcal{N}} \int dN_{g,1}^C \int dN_{g,2}^C \ldots \int dN_{g,d}^C \frac{1}{\widehat{H}_d - \omega + i0^+}\,,
\end{equation}
because even if the spectrum of $\widehat{H}_d$ is independent on
$N_{gj}^C$, the eigenbasis of $\widehat{H}_d$ is not. Only after performing the trace,
\begin{equation}
    \operatorname{tr}\left[\frac{1}{\widehat{H}_d - \omega + i0^+}\right] = \sum_{n} \frac{1}{\epsilon_{n} - \omega + i0^+}\,,
\end{equation}
we arrive at a quantity that does not depend on the eigenbasis.

We now use the gauge-dependent statement in Eq.~\eqref{eq_gauge_equality} for our advantage. First,
we represent the full Green's function in a Dyson series
\begin{equation}
    G^{R}\left(\omega\right) = \operatorname{tr}\left[\frac{1}{\widehat{H}_{0} - \omega + i0^+} \sum_{k=0}^\infty \left(-\widehat{H}_\text{trap} \frac{1}{\widehat{H}_{0} - \omega + i0^+}\right)^k\right],
\end{equation}
and now we integrate it over all the offset charges. Importantly,
due to the invariance of the spectrum on $N_{g,j}^C$, we can take the integrals with respect to $N_{g,j}^C$ in the limit of diverging bounds (that is, for each $N_{g,j}^C$ the lower and upper integral bounds approach $-\infty$
and $+\infty$, respectively). Therefore, in each order $k$ of the Dyson series, the
dominant term is the term of highest order in $N_{g,j}^C$ (which is an even power of $N_{g,j}^C$, such that the integral does not vanish). Consequently, for the integrated quantity,
we find
\begin{equation}
    G^{R}\left(\omega\right) = \frac{1}{\mathcal{N}} \int dN_{g,1}^C \int dN_{g,2}^C \ldots \int dN_{g,d}^C \operatorname{tr}\left[\frac{1}{\widehat{H}_{0} - \omega + i0^+} \sum_{k=0}^{\infty} \left(-h_\text{trap} \frac{1}{\widehat{H}_{0} - \omega + i0^+}\right)^{k}\right],
\end{equation}
where $h_\text{trap}$ is simply the scalar
\begin{equation}
    h_\text{trap}\left(N_{g,1}^C, N_{g,2}^C, \ldots, N_{g,d}^C\right) = 2E_{C} \sum_{j=1}^d \left(N_{g,j}^C\right)^2\,.
\end{equation}
Consequently, we get
\begin{equation}
    G^{R}\left(\omega\right) = \frac{1}{\mathcal{N}} \int dN_{g,1}^C \int dN_{g,2}^C \ldots \int dN_{g,d}^C\, G^{R}_0 \left[\omega + h_\text{trap} \left(N_{g,1}^C, N_{g,2}^C, \ldots, N_{g,d}^C\right)\right],
\end{equation}
and analogously for $\rho$. Note that while it might seem that the
above statement is equivalent to saying that the spectrum of $\widehat{H}_{0}$
is equal to the spectrum of $\widehat{H}_d$, up to a $N_{g,j}^C$-dependent shift,
this is under no circumstances what happens. The traps distort the
spectrum of $\widehat{H}_d$ in a highly nontrivial, and nonperturbative way, while the spectrum stays independent of $N_{g,j}^C$.
The above identity only holds exactly under the precise series of
steps taken above, that is, only for quantities involving the trace,
and only when fixing the gauge of $\widehat{H}_d$ as prescribed. To conclude, we see that the full Green's function $G^R(\omega)$ can be
invoked by $G^R_0(\omega)$ through a convolution with the well-known density
of states of a parabolic spectrum in $d$ dimensions. For $\rho$, the same relationship holds since $\rho$ follows from the imaginary part of $G^R$, see Eq.~\eqref{eq_GR_and_rho}. 
We thus arrive at Eq.~(4) 
in the main text.

\end{widetext}

\section{Linear response }

The density of states of a quantum circuit can be measured via a current response to an external flux. To that end we consider a small time-dependent external magnetic flux driving the phase \(\varphi_{k} \rightarrow \varphi_{k} + \varphi_{\text{ext}}^{k}\left(t\right) \theta\left(t-t_{0}\right)\). The standard linear response of the current (in the interaction picture) yields
\begin{equation}
	\left\langle \widehat{I}_{j} \left(t\right) \right\rangle_{t} = \left\langle \widehat{I}_{j} \left(t\right) \right\rangle_{0} + \int_{t_{0}}^{t} dt_{1}\, \varphi_{\text{ext}}^{k} \left(t_{1}\right) \chi_{jk} \left(t-t_{1}\right),
\end{equation}
with the current operator \(\widehat{I}_{j} = -2e\, i\bigl[\widehat{H}_{J}, \widehat{N}_{j}\bigr]\) and the response function \(\chi_{jj^{\prime}} (t - t_{1}) = i\bigl\langle \bigl[\widehat{I}_{j} (t), \widehat{I}_{j^{\prime}} (t_{1})\bigr]\bigr\rangle_{0} / 2e\). The brackets \(\langle \ldots \rangle_{0}\) denote the expectation value with respect to the unperturbed ground state. Note that the current operator itself in principle also obtains a correction due to the driving. However, this term is trivial since it contributes equally to all frequencies of the response.

Considering the Fourier transformation of the response function, \(\chi_{jj^{\prime}} \left(\omega\right) = i\int_{0}^{\infty} d\tau\, e^{i\omega \tau} \chi_{jj^{\prime}} \left(\tau\right)\),
we can connect the imaginary part of the auto-correlator to the DOS \(\rho(E)\) via
\begin{align}
	\operatorname{Im} \left[\chi_{jj} \left(\omega>0\right)\right] = \pi \left|\left\langle \epsilon_{0}\right| \widehat{I}_{j} \left|\epsilon_{0} + \omega\right\rangle \right|^{2} \rho\left(\epsilon_{0} + \omega\right),
\end{align}
with eigenenergies \(\epsilon_{n}\) and eigenstates \(|\epsilon_{n}\rangle\); see Eq.~(5) 
of the main text.


\end{document}